\documentclass[aps,prd,preprint,groupedaddress]{revtex4-1}
\usepackage{hyperref}
\usepackage{epsfig}
\usepackage{multirow}

\newcommand{\gev}{\ \rm GeV}

\newcommand{\kshort}{K^{0}_{s}}
\newcommand{\piz}{\pi^{0}}
\newcommand{\dzometa}{ D^{0} \rightarrow \omega \eta}
\newcommand{\theresult}{(1.78 \pm 0.19 \pm 0.14)\times 10^{-3}}
\newcommand{\mom}{\gev/c}
\newcommand{\mass}{\gev/c^{2}}

\begin{document}

\preprint{WSU-HEP-1804}

\title{Measurement of $D^{0} \rightarrow \omega \eta$ Branching Fraction with CLEO-c Data}


\author{M. Smith}
\affiliation{Lawrence Technological University}
\author{D. Cinabro}
\affiliation{Wayne State University, Department of Physics and Astronomy}
\email{david.cinabro@wayne.edu}
\author{P. Naik}
\affiliation{University of Bristol}


\date{July 5, 2018}

\begin{abstract} Using CLEO-c data, we confirm the observation of $ \dzometa $ by BESIII.
In the Dalitz Plot of $ D^{0} \rightarrow \kshort \eta \piz $, we find 
a background in the $ \kshort(\rightarrow \pi^{+} \pi^{-}) \piz$ projection with a 
$m(\pi^{+} \pi^{-} \piz)$ equal to the $ \omega(782) $ mass.  In a direct search for 
$ \dzometa $ we find a clear signal and measure $ \mathcal{BF}_{\dzometa} = \theresult $,
in good agreement with BESIII.
\end{abstract}


\maketitle

The recent observation by BESIII of $\dzometa$~\cite{besiii} gave clarity to us of
a mystery we noted in CLEO-c data.
In the Dalitz Plot of $ D^{0} \rightarrow \kshort \eta \piz $, we observed
an anomalous peak at 0.6 $({\rm GeV}/c^2)^2$ in the $m(K^{0}_{s} \piz)^2$ fit projection.  
The BESIII observation leads us to think that this peak is due to 
an $\omega(782) \rightarrow \pi^{+} \pi^{-} \piz $ candidate whose charged 
pions are mis-reconstructed as a $\kshort$.  This decay channel has been 
predicted to have a $\mathcal{BF} = (3.3 \pm 0.2)\times 10^{-3}$\cite{rosner10}.  
Charge conjugation is implied throughout.  Since the decay can proceed from both
a $D^0$ and a $\bar D^0$ and we do no additional reconstruction to find 
the $D$ flavor, we are actually measuring the average of the branching 
fractions of $D^0 \to \omega \eta$ and
$\bar D^0 \to \omega \eta$.

The CLEO-c detector and its experimental methods have been
described in detail elsewhere~\cite{cleo-c}.
This analysis was performed on 818 pb${}^{-1}$
of $e^+e^- \to \psi(3770)$ data with center-of-mass energy $E_{\rm cm} = 3.774$ GeV. 
All $ D^{0}/\bar D^{0} $ candidates are reconstructed 
from $\pi^{\pm}$, $\piz$, and $\eta$ that pass standard selection 
criteria described elsewhere\cite{bigdhadron}.  Charged tracks 
must be well reconstructed and pass basic track quality selections.  
We require a track momentum between $0.050\mom \leq p \leq 2\mom$ 
and the tracks consistent with coming from the interaction region.  
We use the specific ionization, $dE/dx $, 
from the drift chambers and the Ring Imaging CHerenkov (RICH) detector 
to identify our selected tracks as $ \pi^{\pm} $.  
If $dE/dx $ is valid, we require a three standard deviation consistency 
with the $ \pi^{\pm} $ hypothesis.  For tracks with $p \geq 0.70\gev$ 
and $|\cos\theta| < 0.8 $ we can use RICH information as well.  If both 
RICH and $dE/dx $ are valid, we require the 
combined log-likelihood $ \mathcal{L}_{\pi K} \leq 0$ where
\begin{equation}
\mathcal{L}_{\pi K} = \sigma^{2}_{\pi}-\sigma^{2}_{K} + L_{\pi}-L_{K}
\end{equation}
with $L_{h}$ is the log-likelihood of the hypothesis from the RICH information.


We reconstruct $ \pi^{0} $ and $ \eta $ candidates as $ \mathrm{neutral} \rightarrow \gamma \gamma $.
The unconstrained mass is calculated under the assumption 
that the photons originate from the interaction point.  We require this mass to be within 
$ 3\sigma $ of the nominal $ \pi^{0}/\eta $ mass.  A subsequent kinematic fit 
must not be obviously bad, $ \chi^{2} < 10000 $.  We reject neutral 
candidates with both photons detected in the endcap of our calorimeter and 
explicitly reject any photon showers with a matched track.  
Aside from mass values the selections are identical for $\pi^0$ and $\eta$ candidates.

We reconstruct $ D^{0} $ candidates from $\pi^{+}\pi^{-}\piz\eta $ combinations. 
We make an initial requirement that the invariant mass 
$m(\pi^{+}\pi^{-}\piz\eta)$ be within 0.100$\gev/c^{2}$ of the Particle Data Group PDG~\cite{pdg} 
average $D^{0}$ mass.  
We select $\omega(782)$ candidates with the $\pi^{+}\pi^{-}\piz$ invariant mass, $m_\omega$.
We choose selections on $m_\omega$, the beam-constrained mass of $\omega\eta$ ($ \mathrm{M_{bc}} $),
and their $\Delta $E in an iterative procedure making selections on two of the three, 
fitting a Gaussian signal plus smooth backgrounds in the third, making a selection based on 
the fit results, and repeating until the selection values do not change.  In $\mathrm{M_{bc}}$ we 
fit the background to an Argus function, and use a $\mathrm{4^{th}}$ order polynomial 
in $ \Delta $E and $ m_\omega $.  Unlike for $\mathrm{M_{bc}}$ there is 
no physics motivated background shape for $\Delta $E and $m_\omega $, and 
we chose the polynomial of high enough order 
to give a reasonable model of background without adding meaningless nuisance parameters. 
We use the signal mean and standard deviation from one fit to make three standard deviation selections 
on the other plots.  We generate 50000 simulated signal $D^{0}/ \bar D^{0}$ events 
to measure the efficiency of our reconstruction and to determine the optimal 
widths to use in fitting to the data.   We take the yield from $ \mathrm{M_{bc}} $ 
and $ \Delta $E
as  our measurements of the $ D^{0} $ yield in the simulation.  From the 
value of $ \mathrm{M_{bc}} $ yield, we find an efficiency of $ (17.49 \pm 0.216)\% $.

\begin{figure}[ht]
\begin{center}
\epsfig{file=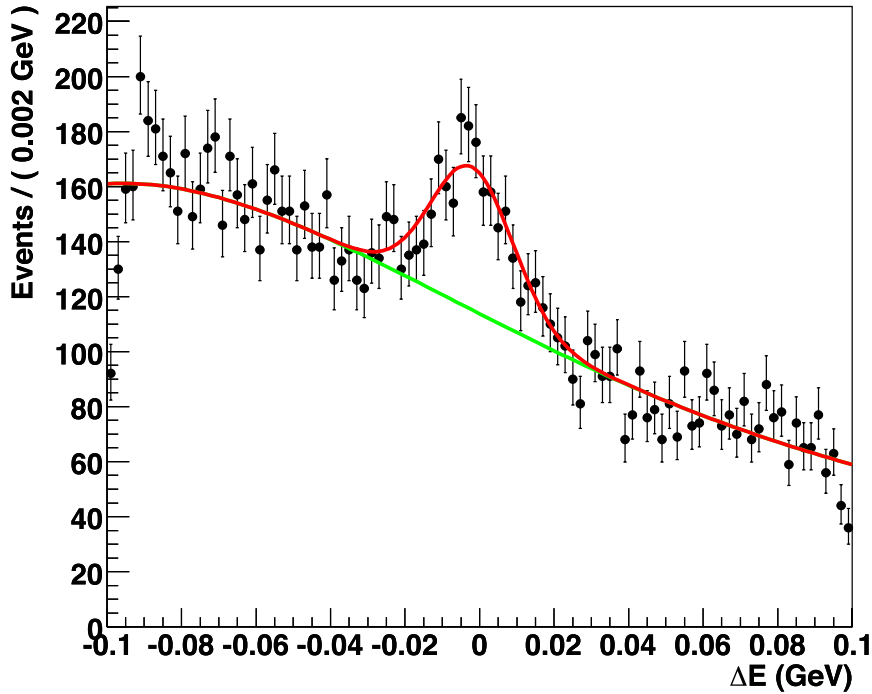,width=5in}
\caption{The $ \Delta $E distribution and fit described in text after three standard deviation signal 
selections for the $ \omega(782) $ and on $\mathrm{M_{bc}}$. }
\label{fig:sigde}
\end{center}
\end{figure}

\begin{figure}[ht]
\begin{center}
\epsfig{file=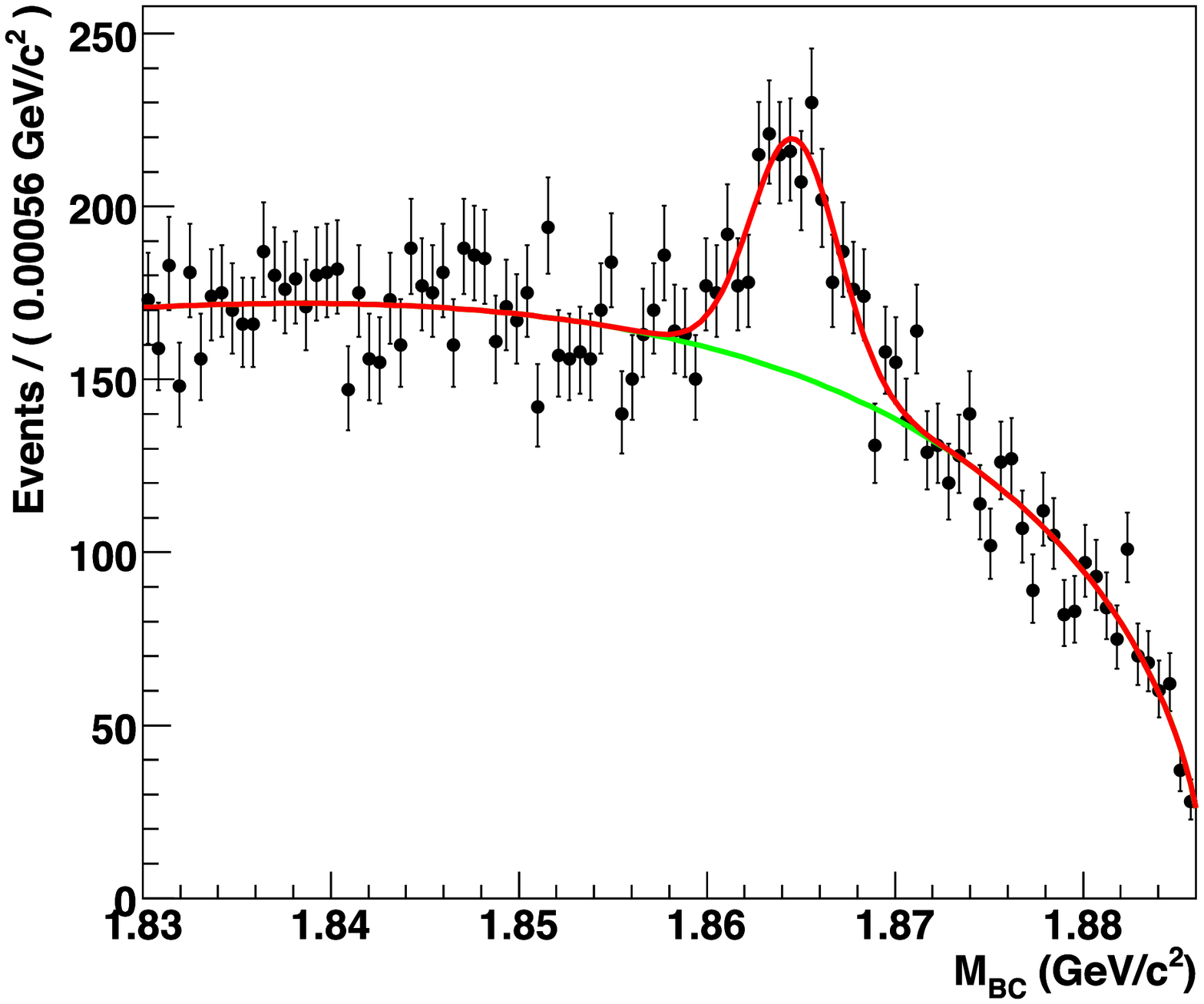,width=5in}
\caption{The $\mathrm{M_{bc}}$ distribution and fit described in text after the
three standard deviation signal selections on $ \omega(782) $ and $ \Delta $E.  }
\label{fig:sigmbc}
\end{center}
\end{figure}

\begin{figure}[ht]
\begin{center}
\epsfig{file=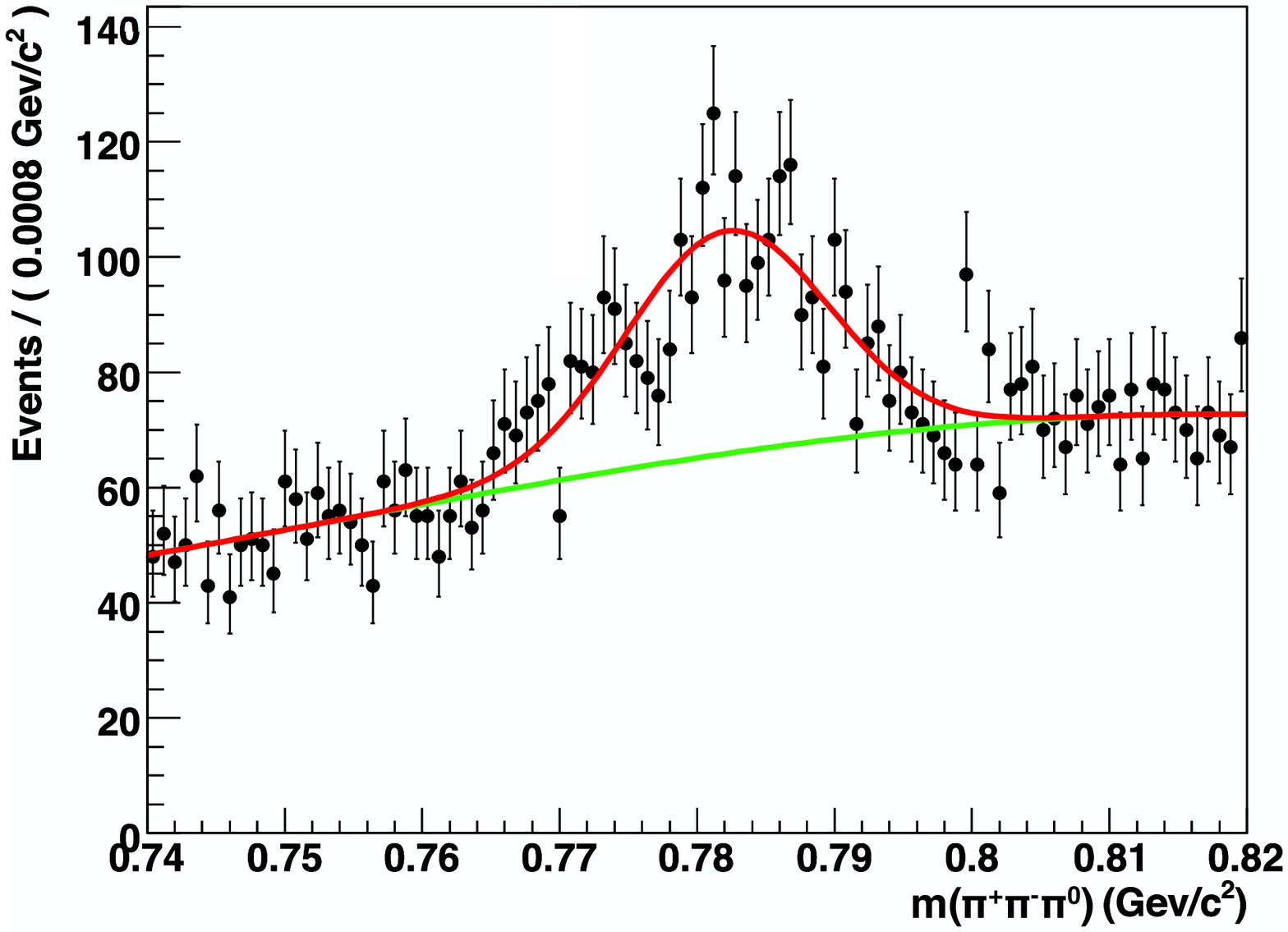,width=5in}
\caption{The $ m(\pi^{+}\pi^{-}\piz) $ invariant mass distribution after 
signal selections in $ \Delta $E and $\mathrm{M_{bc}}$.  The displayed fit
is used to determine $\omega$ candidate selection as described in the text.}
\label{fig:sigomega}
\end{center}
\end{figure}

The same process is performed in data, but with the widths obtained 
in signal simulation fixed in fits to the data distributions.
We choose $ \omega(782) $ candidates which have $ 0.76016\mass \leq m(\pi^{+}\pi^{-}\piz) \leq 0.80432\mass $.
The $m(\pi^{+}\pi^{-}\piz) $ mass fit is used to 
select $ \omega(782) $ candidates, but not for measurement of the $ D^{0} $ yield.  
The $\Delta$E, $ E_{\pi^{+}\pi^{-}\piz\eta} - E_{Beam} $, distribution
is shown in Figure \ref{fig:sigde}.  We set this selection to
$-0.03525 \gev \leq \Delta E \leq 0.03117\gev$.
The beam-constrained mass, $ \mathrm{M_{bc}}^{2}c^4 \equiv E^{2}_{Beam} - p^{2}c^2 $, distribution and fit is 
shown in Figure \ref{fig:sigmbc}, and we select $ 1.857675\mass \leq \mathrm{M_{bc}} \leq 1.871685\mass  $.
The $ \mathrm{M_{bc}} $ and $ \Delta $E fit yields can both be used as measurements of the $ \dzometa $ yield.  
Raw signal yields are $711 \pm 65$ from the $\mathrm{M_{bc}}$ fit
and $720 \pm 70$ from the $\Delta $E fit.
We show the $ m(\pi^{+}\pi^{-}\piz) $ invariant mass distribution after the
selections on $\mathrm{M_{bc}} $ and $ \Delta $E in Figure~\ref{fig:sigomega},
noting that there is a clear $\omega$ signal.

Above, we assume the $\omega(782)$ is strongly related to the reconstruction of the $D^{0}$ 
and its $ \mathrm{M_{bc}} $.  To better visualize this relation, we observe the two dimensional
plot of $\omega(782)$ mass versus $\mathrm{M_{bc}}$, subject to a three standard deviation $\Delta$E cut.
We clearly see a well-populated region near 
the intersection of the $D^{0}$ $ \mathrm{M_{bc}} $ and $\omega(782)$ mass
rising above the large background.  We 
also fit the $ \mathrm{M_{bc}} $ distribution below and above the $\omega(782)$ selections. 
We find no clear $D^{0}$ signal presence in these sidebands.

\begin{figure}[ht]
\begin{center}
\epsfig{file=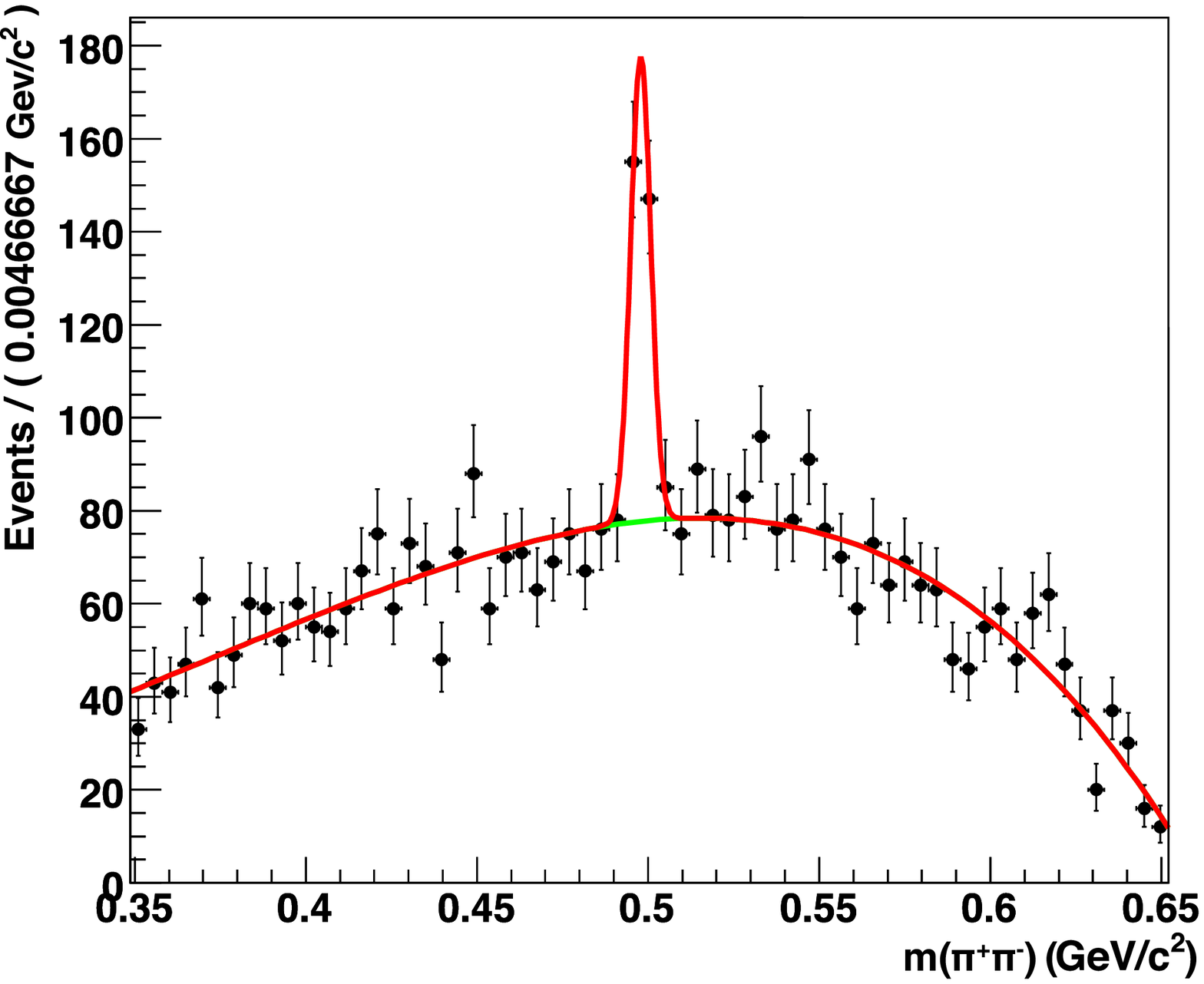, width=5in}
\caption{The $ \mathrm{m}(\pi^{+} \pi^{-}) $ distribution and fit described in text for the $\omega(782)$ candidates
as described in the text. }
\label{fig:kshortfit}
\end{center}
\end{figure}

We expect there to be some $\kshort$ contamination in our $\omega(782)$ signal; 
after all we began with the opposite in $\kshort \eta \piz$.  For our signal 
candidates we show the $ \mathrm{m}(\pi^{+} \pi^{-}) $ distribution in 
Figure \ref{fig:kshortfit}.  
There is a clear $ \kshort $ peak.  This is fit using a Gaussian  ``signal" 
and $\mathrm{4^{th}}$ order polynomial ``background" using the signal region selections above.
We use this to estimate the $ \kshort$ background.
We subtract the ``signal" yield in 
Figure \ref{fig:kshortfit} from our previous results.  We determine 
how many of the $ 158 \pm 20 $ $ \kshort $ events should be subtracted 
by examining $\mathrm{M_{bc}}$ in three regions: three standard deviations around 
the $\kshort$ mean and the two sidebands.  We fit $\mathrm{M_{bc}}$ 
using the previously outlined method, and find the signal and background 
yields under the peak.  Using the signal fraction in the $ \kshort $ region, 
we subtract $ 43 \pm 17 $ from the observed yields.  
The $ \kshort $ subtraction value includes a $ 10\% $ uncertainty 
due to our inability to precisely know how many  $ \kshort $ are
``signal'' versus ``background.''

\begin{table}[ht]
\caption{Signal and Background Yields From $\mathrm{M_{bc}}$, Comparing Three $ \mathrm{m}(\pi^{+} \pi^{-}) $ Regions}
\medskip
\begin{center}
\begin{tabular}{llll}
\hline
\hline
& \multicolumn{3}{c}{In Relation to $ \kshort $ Peak} \\
\hline
$ 3\sigma ~ \mathrm{M_{bc}} $ & Below &  In & Above   \\
\hline
Signal      & 347         & 122         & 229         \\
Background  & 1749        & 327         & 1649        \\
Sig/Total   & $ 16.56\% $ & $ 27.17\% $ & $ 12.19\% $ \\
\hline
\hline
\end{tabular}
\end{center}
\label{tab:kshortcompare}
\end{table}

In a second method of accounting for $ \kshort $ contamination 
we veto the $ \kshort \piz $ contribution to $\omega(782)$ by removing
the $\kshort$ region in $ \mathrm{m}(\pi^{+} \pi^{-}) $.  Aside from the veto, 
the analysis is identical to that described above.  We determine a new efficiency in 
fits to the veto $\mathrm{M_{bc}}$ distribution of
$ (16.13 \pm 0.208)\% $ which represents an $ 7.8\% $ reduction with 
respect to the efficiency without the $\kshort$ veto.

Repeating the data analysis with the $ \kshort $ veto,
Table \ref{tab:sigselections_ks} contains the $ \kshort $ veto analysis yields.
Table \ref{tab:sigyields} contains the yields from $\Delta $E and $\mathrm{M_{bc}}$ 
corrected by both $ \kshort $ subtraction and veto, as well as their associated 
efficiencies and efficiency corrected yields.  

\begin{table}[ht]
\caption{Summary of Signal Selections with $ \kshort $ Veto}
\medskip
\begin{center}
\begin{tabular}{c}
\hline
\hline
Signal Selections \\
\hline
$ m(\pi^{+} \pi^{-}) \leq 0.48902\mass ~ \mathrm{or} ~ \mathrm{m}(\pi^{+} \pi^{-}) \geq 0.50672\mass $\\
$ 0.76010\mass  \leq m(\pi^{+}\pi^{-}\piz) \leq 0.80474\mass $ \\
$ -0.03551\gev \leq \Delta E              \leq 0.03145\gev $ \\
$ 1.857738\mass \leq \mathrm{M_{bc}}       \leq 1.871802\mass  $ \\
\hline
\hline
\end{tabular}
\end{center}
\label{tab:sigselections_ks}
\end{table}

\begin{table}[ht]
\caption{Signal Yields from Fittings Accounting for $ \kshort $ Effects}
\medskip
\begin{center}
\begin{tabular}{llclc}
\hline
\hline
& Type & Signal Yield & Signal Efficiency & Yield/Efficiency\\
\hline
\multirow{2}{*}{$ \kshort $ Events Subtracted} & $\mathrm{M_{bc}}$  & $ 667 \pm 67 $ & $ (17.49 \pm 0.22)\% $ & 3819\\
											   & $ \Delta $E        & $ 677 \pm 72 $ & $ (17.06 \pm 0.22)\% $ & 3969\\
\multirow{2}{*}{$ \kshort $ Veto}   	       & $\mathrm{M_{bc}}$  & $ 596 \pm 62 $ & $ (16.13 \pm 0.21)\% $ & 3694\\
											   & $ \Delta $E        & $ 597 \pm 67 $ & $ (15.79 \pm 0.21)\% $ & 3780\\
\hline
\hline
\end{tabular}
\end{center}
\label{tab:sigyields}
\end{table}

The analyses described above used widths from Signal Monte Carlo fixed in the data fits.  
When we float the data widths in the $\kshort$ veto analysis, 
we find $637 \pm 89$ and $521 \pm 85$ for the $\mathrm{M_{bc}}$ 
and $ \Delta $E signal yields, respectively.  These values greatly differ from those with 
fixed widths, and indeed greatly from each other.  We will use the difference between 
fixed and floating $\mathrm{M_{bc}}$ yields as a systematic uncertainty.
 
We calculate the Branching Fraction using
\begin{equation} \label{eq:bf} \mathcal{BF} = \frac{N_{\dzometa}}{2 \epsilon_{\dzometa} N_{ D^{0} \bar D^{0} } \mathcal{BF}_{\omega \rightarrow \pi^{+} \pi^{-} \piz } \mathcal{BF}_{\eta \rightarrow \gamma \gamma} \mathcal{BF}_{\piz \rightarrow \gamma \gamma} } \end{equation}
where $ N_{\dzometa} $ is the observed yield and $ N_{ D^{0} \bar D^{0} }$ is the total number
of $ D^{0} / \bar D^{0} $ events.  We calculate $ N_{ D^{0} \bar D^{0} } $ by multiplying 
$\sigma( e^{+}e^{-} \rightarrow D^{0} \bar D^{0}) $ previously 
reported by CLEO\cite{bigdhadron} and our integrated luminosity.  
Table \ref{tab:ext_input} contains the Branching Fraction inputs.

Comparing the Yield/Efficiency results in Table \ref{tab:sigyields} we see the 
$ \kshort $ Subtraction and Veto are both acceptable methods to deal with 
$ \kshort $ contamination giving consistent results.  The four efficiency corrected
yields have a standard deviation of 115, which is $3.0\%$ of the 
average efficiency corrected yield of 3816.  The efficiency corrected yields 
are larger in the subtraction method and this method has a 
conceptual problem.  Our subtraction choice is a best guess;  
there is no clear way to determine how many $ \kshort $ actually 
contribute to the signal rather then the background.  

We therefore take the $\mathrm{M_{bc}}$ yield from 
the $ \kshort $ veto analysis as the best measurement.
Comparing using $\mathrm{M_{bc}}$ and $ \Delta $E to extract the yield, 
we have a fortunately small $ \pm 1 $ systematic uncertainty from the difference in signal 
yield and $\pm 0.34\%$ uncertainty from the difference in Efficiency.  These
give a $2.13\% $ relative uncertainty on the efficiency corrected yield.  
We also have $\pm 41$ systematic on the yield
due to the difference between using fixed and floating widths in 
$\mathrm{M_{bc}}$ fits.  These two yield uncertainties give us a total systematic
uncertainty on the yield. 
We find $ \mathcal{BF}_{\dzometa} = \theresult $.  The statistical 
uncertainty comes from the statistical uncertainty in the 
signal yield.  All of the uncertainties are summarized in 
Table \ref{tab:bfcompare}.  The contribution 
from $\mathcal{BF}(\piz \rightarrow \gamma \gamma)$ is negligible.

\begin{table}[ht]
\caption{Summary of Branching Fraction Inputs. Branching Fractions are PDG\cite{pdg} values. Uncertainties 
are statistical and systematic, respectively. }
\medskip
\begin{center}
\begin{tabular}{ll}
\hline
\hline
Quantity        & Value \\
\hline
 Signal Yield 												 & $  596 \pm 62 \pm 1 $ 				  \\
 Efficiency 												 & $ (16.13 \pm 0.208 \pm 0.34)\% $		  \\
$\mathcal{BF}(\omega(782) \rightarrow \pi^{+} \pi^{-} \piz)$ & $ (89.2 \pm 0.7)\%$ 			  \\
$\mathcal{BF}(\eta \rightarrow \gamma \gamma)$      		 & $ (39.31 \pm 0.2)\% $ 		  \\
$\mathcal{BF}(\piz \rightarrow \gamma \gamma)$ 				 & $ (98.823 \pm 0.034)\% $ 	  \\
$\sigma( e^{+}e^{-} \rightarrow D^{0} \bar D^{0}) $ 		 & $ (3.66 \pm 0.03 \pm 0.06)nb $ \\
Luminosity      											 & $ 818 \pm 8 \mathrm{pb}^{-1} $ \\
\hline
$N_{ D^{0} \bar D^{0} }$ & 2993880 \\
\hline
\hline
\end{tabular}
\end{center}
\label{tab:ext_input}
\end{table}

\begin{table}[ht]
\caption{Summary of the uncertainties on $ \mathcal{BF}_{\dzometa}$.}
\begin{center}
\begin{tabular}{llll}
\hline
\hline
Source             											 & Value ($\times 10^{-3}$) \\
\hline
Statistical on Yield        								 & $\pm 0.19 $ \\
\hline
 Signal Yield        										 & $\pm 0.125$ \\
 MC Efficiency 			       								 & $\pm 0.038$ \\
Luminosity          										 & $\pm 0.0178$ \\
Cross Section          										 & $\pm 0.0326$ \\
$\mathcal{BF}(\omega(782) \rightarrow \pi^{+} \pi^{-} \piz)$ & $\pm 0.0140$ \\
$\mathcal{BF}(\eta \rightarrow \gamma \gamma)$ 				 & $\pm 0.00906$ \\
\hline
Total Systematic    										 & $\pm 0.137$ \\
\hline
Total Uncertainty    										 & $\pm 0.23$ \\
\hline
\hline
\end{tabular}
\end{center}
\label{tab:bfcompare}
\end{table}

In summary, in the CLEO-c data we have observed $D^{0} \rightarrow \omega \eta$ and measure
the average of $D^0 \to \omega \eta$ and ${\bar D}^0 \to \omega \eta$ as
\begin{equation}
 \mathcal{BF}(\dzometa) = \theresult.
\end{equation} 
This agrees with the previous observation by BESIII.
Our measured branching fraction is roughly a factor of two smaller
than predicted.  We note that this $D^0$ decay mode is a CP-eigenstate making it a potentially
valuable tool in heavy flavor analysis.

This investigation was done using CLEO data, and as members of the former CLEO 
Collaboration we thank it for this privilege. This research was supported 
by the U.S.~National Science Foundation.

\end{document}